\documentclass[reprint,aps,prl,superscriptaddress,longbibliography]{revtex4-1}
\usepackage[utf8]{inputenc}
\usepackage{color}
\usepackage{times}
\usepackage{amsmath}
\usepackage{amssymb}
\usepackage{stmaryrd}
\usepackage{makecell}
\usepackage[caption=false]{subfig}
\usepackage{xr-hyper}
\usepackage{graphicx}
\usepackage{tikz-cd}
\usepackage{adjustbox}
\usepackage[unicode=true,
 bookmarks=true,bookmarksnumbered=false,bookmarksopen=false,
 breaklinks=false,pdfborder={0 0 1},backref=false,colorlinks=true]
 {hyperref}
\hypersetup{
 linkcolor=magenta, urlcolor=blue, citecolor=blue, pdfstartview={FitH}, hyperfootnotes=true, unicode=true}
\makeatletter
\@ifundefined{textcolor}{}
{%
 \definecolor{BLACK}{gray}{0}
 \definecolor{WHITE}{gray}{1}
 \definecolor{RED}{rgb}{1,0,0}
 \definecolor{GREEN}{rgb}{0,1,0}
 \definecolor{BLUE}{rgb}{0,0,1}
 \definecolor{CYAN}{cmyk}{1,0,0,0}
 \definecolor{MAGENTA}{cmyk}{0,1,0,0}
 \definecolor{YELLOW}{cmyk}{0,0,1,0}
}

\def\be{\begin{equation}}
\def\ee{\end{equation}}
\def\bc{\begin{center}}
\def\ec{\end{center}}

\def\r2{{\sqrt{2}}}

\def\bea{\begin{eqnarray}}
\def\eea{\end{eqnarray}}

\definecolor{burntorange}{rgb}{0.8, 0.33, 0.0}

\newcommand{\doublewidetilde}[1]{{%
  \mathpalette\double@widetilde{#1}%
}}
\newcommand{\double@widetilde}[2]{%
  \sbox\z@{$\m@th#1\widetilde{#2}$}%
  \ht\z@=.9\ht\z@
  \widetilde{\box\z@}%
}

\usepackage{amsfonts}\usepackage{tabularx}\usepackage{dcolumn}\usepackage{bm}\usepackage{graphicx}\usepackage{epstopdf}

\hypersetup{urlcolor=blue}
\usepackage{xr}
\usepackage{lipsum}
\usepackage{tensor}
\usepackage{braket}

\usepackage[capitalise,compress]{cleveref}
\crefname{section}{Sec.}{Secs.}
\Crefname{section}{Section}{Sections}
\crefrangelabelformat{equation}{\textup{(#3#1#4)}--\textup{(#5#2#6)}}

\makeatother

\usepackage{xcolor}

\definecolor{darkgreen}{rgb}{0.0, 0.6, 0.13}

\usepackage{amsfonts}\usepackage{tabularx}\usepackage{dcolumn}\usepackage{bm}\usepackage{graphicx}\usepackage{epstopdf}

\hypersetup{urlcolor=blue}

\usepackage{xr}
\usepackage{xr-hyper}
\usepackage{hyperref}
\makeatletter
\newcommand*{\addFileDependency}[1]{% argument=file name and extension
  \typeout{(#1)}% latexmk will find this if $recorder=0 (however, in that case, it will ignore #1 if it is a .aux or .pdf file etc and it exists! if it doesn't exist, it will appear in the list of dependents regardless)
  \@addtofilelist{#1}% if you want it to appear in \listfiles, not really necessary and latexmk doesn't use this
  \IfFileExists{#1}{}{\typeout{No file #1.}}% latexmk will find this message if #1 doesn't exist (yet)
}
\makeatother

\makeatother
\begin{document}

%\title{Diffusion map on SPT phase}
\title{Unsupervised Learning of Symmetry Protected Topological Phase Transitions}
\author{En-Jui Kuo}
\email[]{These authors contributed equally to this paper.}
\affiliation{Department of Physics, University of Maryland, College Park, MD 20742, USA}
\affiliation{Joint Quantum Institute, NIST/University of Maryland, College Park, MD 20742, USA}
\affiliation{Joint Center for Quantum Information and Computer Science, University of Maryland, College Park,
Maryland 20742, USA}

\author{Hossein Dehghani}
\email[]{These authors contributed equally to this paper.}
\affiliation{Department of Physics, University of Maryland, College Park, MD 20742, USA}
\affiliation{Joint Quantum Institute, NIST/University of Maryland, College Park, MD 20742, USA}

\begin{abstract}
Symmetry-protected topological (SPT) phases are short-range entangled phases of matter with a non-local order parameter which are preserved under a local symmetry group. Here, by using unsupervised learning algorithm, namely the diffusion maps, we demonstrate that can differentiate between symmetry broken phases and topologically ordered phases, and between non-trivial topological phases in different classes. In particular, we show that the phase transitions associated with these phases can be detected in different bosonic and fermionic models in one dimension. This includes the interacting SSH model, the AKLT model and its variants, and weakly interacting fermionic models. Our approach serves as an inexpensive computational method for  detecting topological phases transitions associated with SPT systems which can be also applied to experimental data obtained from quantum simulators. 

\end{abstract}

\maketitle

{\it Introduction.---}
In the last two decades, theoretical prediction and experimental observation of topological phases of matter \cite{wen2017Colloquium} have been one of the most remarkable advancements in the field of condensed matter physics. In topological quantum phases of matter, although, the energy spectrum is gapped, a Landau-Ginzburg framework based on existence of local order parameters, cannot explain some of the most important features of these system such as the existence of symmetry protected edge states  \cite{wen1990ground}. While initially most efforts were mostly focused on non-interacting topological insulators,  \cite{Kane2005Topological, Bernevig2006Quantum, Topological2007Moore, Fu2007Topological,  Konig2007Quantum, Hsieh2008Topological, Hasan2011Three} in recent years, attention has turned into interacting topological phases of matter. %including symmetry protected topological phases (SPT) and symmetry enriched topological phases. 

Classifying topological insulators and superconductors which are described by non-interacting fermionic Hamiltonians is less challenging than characterizing interacting topological phases of matter. In particular since the spectrum of non-interacting models are often exactly solvable, classifying them can be done rigorously via several methods including the classification of random matrices \cite{Altland1997Nonstandard}, Dirac operators \cite{Ryu2008Classification} or more abstractly via K-theory \cite{kitaev2009periodic}. Nevertheless, for interacting topological systems, more refined methods such as group cohomology methods have been proposed \cite{chen2012symmetry, Barkeshli2019Symmetry}. Correspondingly, for the former systems, using the wave functions of the Hamiltonian or the Green's functions in the momentum space one can define their corresponding topological invariants, while for the latter, one needs to define non-local order parameters in the real space \cite{Pollmann2012symmetry, Cirac2012order, Shiozaki2017, Dehghani2021Extraction} which is both computationally and experimentally costly to probe \cite{Elbeneaaz3666, Cian2021ManyBody}. Therefore, identification of topological properties in a given interacting topological systems, is computationally more involved and designing efficient computational methods to detect possible topological phase transitions is an indispensable task for studying these phases. 
%are relatively straightforward to be understood and classified. 
%This task can be done rigorously via the classification of random matrices
%\cite{Altland1997Nonstandard}, Dirac operators \cite{Ryu2008Classification} or more abstractly via K-theory \cite{kitaev2009periodic}. Nevertheless, for interacting topological systems, more refined methods such as group cohomology methods have been proposed \cite{chen2012symmetry, Barkeshli2019Symmetry}. Correspondingly, for the former systems, using the wave functions of the Hamiltonian or the Green's functions in the momentum space one can define their corresponding topological invariants, while for the latter, one needs to define more complicated non-local order parameters in the real space \cite{Pollmann2012symmetry, Cirac2012order, Shiozaki2017, Dehghani2021Extraction} which is both computationally and experimentally costly to probe \cite{Elbeneaaz3666, Cian2021ManyBody}. Therefore, identification of topological properties in a given interacting topological systems, is computationally more involved and designing efficient computational methods to detect possible topological phase transitions is an indispensable task for studying these phases. 
 
On the other hand over the past few years machine learning (ML) has emerged as a powerful tool to assist physicists to study a plethora of different problems in condensed matter and quantum sciences. A non-exhaustive list of notable examples include classifying phases of matter \cite{carrasquilla2017machine,wetzel2017unsupervised,van2017learning, kuo2021decoding}, studying non-equilibrium dynamics of physical systems \cite{van2018learning,schindler2017probing,seif2019machine}, simulating dynamics of quantum systems \cite{carleo2017solving, torlai2018neural,carrasquilla2019reconstructing}, and augmenting capabilities of quantum devices \cite{seif2018machine,torlai2019integrating}. In particular in classifying applications, most of such techniques rely on supervised ML techniques where the ML algorithms after being trained with labeled systems learns to classify systems with new parameters \cite{carrasquilla2017machine, DasSarma2017Machine, Kim2018Machine, Zhang2018Machine, seif2019machine, Ni2019Machine, Huang2021Information}. However, more interestingly in unsupervised machine no prior knowledge of the phase of the systems is provided and the algorithm by detecting the hidden structure of the input data learns to cluster wave functions or Hamiltonians in different phases \cite{Wang2016Discovering,Hu2017Discovering, wetzel2017unsupervised}. %A successful approach is using diffusion maps have been shown to be a powerful tool to detect band topology and equilibrium symmetry broken phases. Therefore, using unsupervised machine learning techniques can provide a complementary tool to identify new SPT phases. 

 %In quantum simulators being composed of a finite number of particles/qubits, many of these techniques lose their relevance. 
 
%Symmetric protected topological phase is a short range interaction. The bosonic case is classified by using group cohomology \citep{chen2012symmetry}. Symmetry-protected trivial (SPt) phases of matter are the product-state analogue of symmetryprotected topological (SPt) phases. This means, SPT phases can be adiabatically connected to a product state by some path that preserves the protecting symmetry. Moreover, SPt and SPT phases can be adiabatically connected to each other when interaction terms that break the symmetries protecting the SPT order are added in the Hamiltonian.

{\it Goal.---}
In this work, we use unsupervised learning model (ULM), namely the diffusion maps algorithm (DMA), to detect different topological phase transition associated with symmetry protected interacting topological phases with short-range entanglement also known as symmetry-protected topological (SPT) systems \citep{chen2012symmetry}. The diffusion map algorithm which can capture the hidden geometrical and topological structure of data sets \cite{coifman2005geometric}, recently, has been used to classify phases of matter in the local order of the Ising model \cite{lidiak2020unsupervised}, thermal topological vortex structures and temperature driven phase transitions \cite{rodriguez2019identifying}, and band topology \cite{scheurer2020unsupervised}. 

Here, by applying this algorithm to different bosonic and fermionic systems in one dimension which host SPT phase, we demonstrate that this technique can approximately identify topological phase transitions and reproduce the phase diagram of models which in addition to symmetry broken phases can host one or several interacting SPT phases. Hence, in systems where based on physical symmetry arguments, it is known that there are finite possibilities for formation of symmetry broken phases, detection of additional phases via our method is an evidence that the system can host non-trivial SPT phases. We should highlight that since we only need snapshots of the ground state wave functions of the system, unlike other methods which to detect topological phases rely on non-local (string-like) operators \cite{denNijs1989Preroughening, Oshikawa1992Hidden, Hidden1992Kennedy, Levin2006Detecting, Pollmann2012symmetry}, or nonlinear functions of the wave functions such as entanglement entropy and entanglement spectrum of the wave function \cite{Kitaev2006Topological, Haldane2008Entanglement, Calabrese2008Entanglement}, our approach is computationally more convenient.
We also note that %in contrast to previous works which use thermal distributions and thermal sampling to study interacting topological phases transitions \cite{rodriguez2019identifying}, here, 
since we use ground states directly, this approach is also suitable for experimental data %our method serves as an inexpensive technique which can be conveniently applied to experimental data 
obtained from quantum simulators \cite{de2018experimental} and noisy intermediate-scale quantum (NISQ) devices \cite{preskill2018quantum}.  

\section{Diffusion Map}
The DMA is based upon the classical idea of integration that the global structure of a manifold can be determined by traversing it provided that local transition rules to move from one point to another is at hand. To do so, we create a Markov matrix of transitions for the points in the input data set, which guides a random walker to traverse the data set through a diffusion process which results in gaining information about the global geometrical properties of the input data. However, in order to probe the structure of the input data at different scales, a family of transition matrices are employed, and hence the name diffusion ``maps".

More concretely, suppose we have $n$ input data, each represented by a $N$-dimensional vector $\vec{x}_{\alpha}=(x_{\alpha}^{1}, \cdots, x_{\alpha}^{N})$ in $ \mathbb{R}^{N}$ where Greek letters denote the sample indices. For every two sampled vectors $\vec{x}_{\alpha}, \vec{x}_{\beta}$, we use the Euclidean metric to define their distance. To set the local rules of transition between two point we start by defining a symmetric positivity preserving function which for most applications could be a Gaussian kernel:
\be
{\displaystyle k(\vec{x}_{\alpha}, \vec{x}_{\beta})=\exp \left(-{\frac {|\vec{x}_{\alpha}-\vec{x}_{\beta}|^{2}}{\epsilon }}\right)},
\ee
where $\epsilon$ is the scaling hyperparameter of the DMA which determines the speed of the diffusion process of the corresponding random walker and correspondingly the number of clusters.  
We also define the diffusion matrix $K$ which is a version of the graph Laplacian matrix with components $K_{\alpha \beta}=k(\vec{x}_{\alpha},\vec{x}_{\beta}).$ To interpret this matrix as a probability distribution we need to normalize it by the diagonal matrix $D$ whose components are given by ${\displaystyle D_{\alpha \alpha}=d_{\alpha}}$ where $d_{\alpha}=\sum_{\beta}K_{\alpha \beta}$ and plays the role of the local degree of the graph. Then, the normalized Laplacian matrix is defined by 
%Equivalently, one can write this equation in a matrix form by defining $D$ as a diagonal matrix with ${\displaystyle D_{\alpha \alpha}=d_{\alpha}}$. We then compute the graph Laplacian normalization of this new kernel \citep{coifman2006diffusion}:
\be
{\displaystyle M={D}^{-1}K,\,}
\ee
Physically, the matrix $M$ denotes the transition probabilities between different samples such that the probability of transition from sample $\alpha$ to $\beta$ in $t$ time steps is given by the $t$th power of this matrix i.e. $M^t_{\alpha \beta}$. We also notice that $M$ is a stochastic matrix \cite{de2008introduction} with positive entries where each column sums to $1$ and hence, the largest eigenvalue is $1$. 

Next, we compute the diagonal representation of $M$. As in principal component analysis (PCA) where we only keep the largest eigenvalues to determine the number of clusters  \cite{jolliffe2016principal}, here, we also keep the largest eigenvalues in the diagonal representation of $M$. We note that due to the spectral decay of the eigenvalues only a finite number of terms are necessary to achieve a given relative accuracy in reproducing $M$ via its diagonal representation. Therefore, to reach an accuracy labeled by $\delta$, we only need to keep the eigenvalues larger than $1-\delta$ in the spectrum. Since $\delta$ is arbitrary, we can fix it, and then study how the number of relevant clusters vary with the scale parameter $\epsilon$ to probe the underlying geometric structure of the  at different scales. We note that the main advantage of this method over more conventional ULMs such as PCA is that while maintaining the local structures of the data sets, it can also recognize their underlying nonlinear manifolds due to its nonlinear kernel which makes it advantageous compared to other unsupervised learning methods \cite{supp}. %These two features makes this algorithm successful in detecting complicated phases of matter realized by local Hamiltonian which are not detected by other methods such as PCA \cite{supp}. %The steps in applying this algorithm can be summarized as follows:
%\begin{itemize}
%  \item Produce the similarity matrix $K$.
 
%  \item Form the normalized matrix  $M=D^{-1}K$
  
%  \item Compute the eigenvalues of $M$ and plot the phase diagram.
%\end{itemize}

\begin{table*}\label{t1}
\begin{adjustbox}{width=18cm, center}
\resizebox{1.0\textwidth}{!}{
	\begin{tabular}{|c| c| c|c|c|c }\hline
		Hamiltonian    $H$            & parameters  & Type of phase   & $\epsilon$  & $N$  \\\hline
$-J'\sum_{i} \sigma_{2i-1}^{-}\sigma_{2i}^{+}-J\sum_{i} \sigma_{2i}^{-}\sigma_{2i+1}^{+} + h.c$  & $J, J'$ & Topological/Trivial  & $0.088$ & $9$\\\hline
$\sum_{i} \cos(\theta) (\vec{S}_{i}\cdot \vec{S}_{i+1})
+ \sin(\theta) (\vec{S}_{i}\cdot \vec{S}_{i+1})^2$ & $\theta$ & H, Dimer, FM, & $\frac{1}{32}$ & $9$ \\\hline
$\sum_{i}  (\vec{S}_{i}\cdot \vec{S}_{i+1})+ B\sum_{i} S_x^{i}+\sum_{i} D (S_z^{i})^2$ & $B, D$ &  spin polarized, $\mathbb{Z}_2$ SB, H
& $\frac{1}{32}$ & $10$ 	\\\hline
$H=-\sum_{<ij>\sigma}c^\dagger_{i\sigma}c_{j\sigma}-2\Delta_s\sum_jc^\dagger_{j\uparrow}c^\dagger_{j\downarrow} 
 \pm i\Delta_p/2\sum_j(c^\dagger_{j+1\uparrow}c^\dagger_{j\downarrow}+c_{j+1\downarrow}c_{j\uparrow})+ \textrm{h.c.}$
 & $\Delta_s, \Delta_p$  &
 $N=0,\pm 1$ 
 & $\frac{1}{32} $ & $6$
\\\hline
\end{tabular}
}
\end{adjustbox}
\caption{Summary of $1$D Hamiltonians. H labels Haldane phase. Dimerized phase, Dimer phase, ferromagnetic phase, antiferromagnetic phase, XY phase, valence Bound state, $\mathbb{Z}_2$ symmetry breaking (SB) phase, critical phase.}
\end{table*}

\section{Model introduction and results}
\begin{figure}
    \centering
    \includegraphics[scale=0.2]{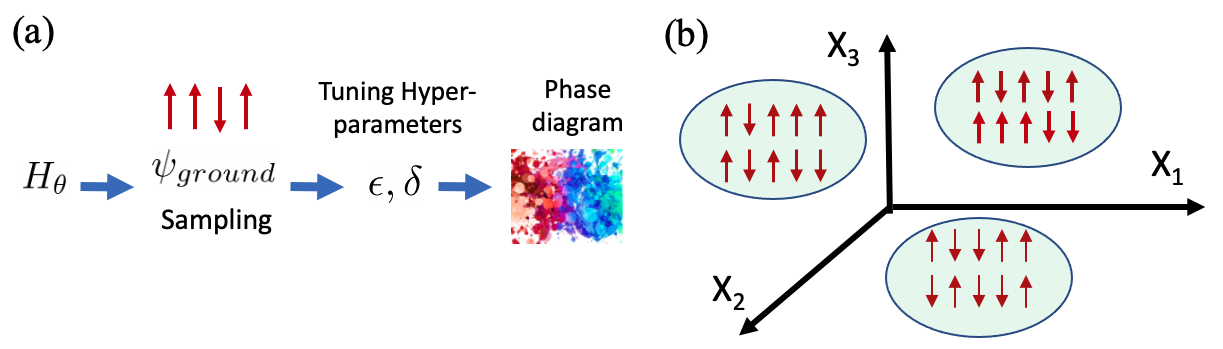}
    \caption{Schematic illustration of the machine-learning algorithms to identify different phases. (a) Flowchart of the diffusion maps algorithm. From a given Hamiltonian, (1) one obtains the ground state, and projectively samples from the ground state, (2) forms the diffusion matrix and tunes the hyperparameters, (3) obtains the phase diagram by finding the conspicuous transitions in the number of eigenvalues. (b) Different ground states in a cluster can transform to each other via a diffusion process in the dataset  basis.}
    \label{fig:illu}
\end{figure}

Let us now apply the DMA explained above to our quantum many-body problem. We imagine we are given a Hamiltonian with a set of unknown parameters $\boldsymbol{\theta} \in \mathbb{R}^s.$ We assume that we have direct access to the ground state of a $N$-particle many-body Hamiltonian and we can sample from its many-body wave function in the configuration space. 
%Such a setting not only frequently occurs in computational studies of may body systems via exact diagonalization, Monte Carlo and tensor network methods, but also is useful for experimental applications with noisy intermediate-scale quantum (NISQ) devices where the state of the system can be projectively measured in a given bases. 
In particular, in this study, for spin-$\sigma$ chains our sampled wave function determines the configuration of the spins in the real space, and therefore, the components of the $\alpha$'th input vector are  $x_{\alpha}^{i}=\{\pm \sigma\}$. For our fermionic and hard-core bosonic Hamiltonians, the measured wave function is represented in the Fock space, and therefore, $x_{\alpha}^i = \{0, 1\}$. In this setting, the schematic flowchart of our classification approach is illustrated in Fig.\ref{fig:illu}, 
%\begin{eqnarray}
%    H_{\boldsymbol{\theta}} \to \psi_{\text{ground}}(\boldsymbol{\theta}) && \to \text{Sampling}  \to \text{Diffusion map}\notag \\ && \to \text{Phase diagram}, 
%\end{eqnarray}
where $ \psi_{\text{ground}}(\boldsymbol{\theta})$ denotes the ground state wave functions. In general, this state is a superposition of different product states. In the next step, the sampling is performed in the space of these product states resulting in the vectors $\vec{x}_{\alpha}(\boldsymbol{\theta})$ by which we can produce the corresponding Laplacian matrix $M(\boldsymbol{\theta})$. %\footnote{For Hamiltonian with ground state degeneracy $\geq 2$ in open boundary conditions, we randomly choose one of the ground states and perform the sampling.}.
Since, across quantum phase transitions usually a change in the relevant degrees of freedom that describe the system occurs, we expect that for a given accuracy $\delta$, once the scale $\epsilon$ is chosen properly, the number of clusters obtained from $M(\boldsymbol{\theta})$ should reveal the underlying phase diagram of the system. %As in Fig.\ref{fig:illu}(b), all states in the same cluster can transform into each other via a diffusion process.

To determine the proper regime for $\epsilon$ as the parameter that describes the dynamics of the diffusion process \cite{coifman2006diffusion}, we use the approximate analysis of the spectrum of $M$ proposed in \cite{ rodriguez2019identifying, lidiak2020unsupervised}. %Considering a random walker at time $t=0$ at the origin of the the data space, the probability distribution of the samples which are covered by the random walker is a function of the Euclidean distance and time which in the continuum limit can be approximated by the heat equation \cite{}. 
Assuming clusters of samples with $r$ particles in different configuration states, the diffusion matrix $K$ will acquire a block-diagonal form with each block corresponding to a single such clusters. In this representation, diagonal elements are close to the identity while off-diagonal components are bounded by $\mathcal{O}(e^{-r/\epsilon})$. Denoting the number of such clusters with $N_c$, after diagonalizing $K$, we will approximately have $N_c$ eigenvalues larger than $1-e^{-r/\epsilon}$ and therefore the accuracy can be approximated by $\delta \sim e^{-r/\epsilon}$. %where $c_1$ is constant of order $\mathcal{O}(1)$.  
%Due to this structure of the spectrum, from a dynamical perspective, up to a time scales $\tau \sim e^{r/\epsilon}$ %\footnote{There is an addition polynomial factor of $\mathcal{O}(r^2$) in the corresponding number of time steps which can be ignored in comparison to the exponential factor given the main text.},
%the diffusion process occurs inside single clusters. Since the diffusion time $\tau$ is arbitrary we fix it by choosing $\delta=10^{-5}.$
Thus by choosing $\epsilon \sim -r/ln \delta$, we can find the number of clusters of a particular size $r$. Consequently, since the number of particles in the clusters can change between $1$ to $N$, we need to span the corresponding range of $\epsilon$, to obtain the best value of $\epsilon$ empirically where changes in $N_c$ show a conspicuous change for different values of $\boldsymbol{\theta}$. Therefore, sharp changes in $N_c$ as a function of $\boldsymbol{\theta}$ can be presumed to trace the topological phase transitions.

In what follows we employ the procedure above, and consider some of the most well-known models which describe SPT phases and we demonstrate that we can reproduce their phase diagram with an acceptable precision. We summarize our models in table \ref{t1} where for all models we have $\delta=0.00001$) and we use periodic boundary conditions.  

%In these models, we verify that DMA can detect the phase diagram by differentiating between symmetry broken and SPT phases. 

\begin{figure}
    \centering
    \includegraphics[scale=0.19]{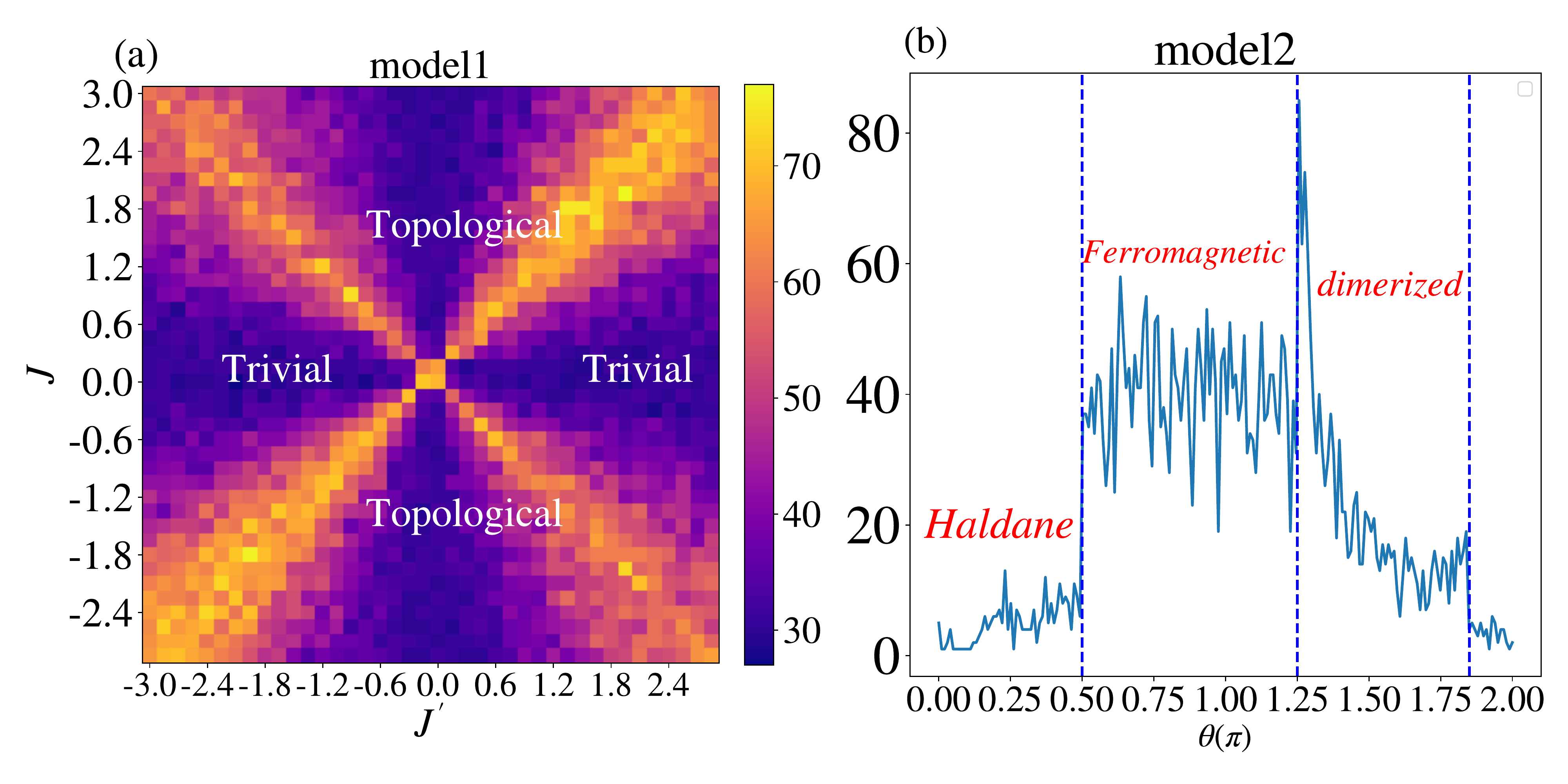}
    \caption{Phase diagrams of model 1 and 2. Colorbar in (a) and vertical axis in (b) represent the number of eigenvalues larger than $1-\delta$ (See text). (a)Model 1 is an interacting bosonic SPT model with a topolgical and trivial phase with phase boundaries at $J=\pm J'$. (b) Model 2 is biquadratic spin chain model whose ground state at $\theta=1/3$ is the AKLT state. The dashed lines illustrate the phase boundaries obtained via DMRG methods.}
    \label{fig:model3}
\end{figure}

{\it Model 1: Nontrivial SPT/Symmetry-protected trivial.---}
In this section, we ask whether we can distinguish between ground states with a trivial phase and a non trivial SPT phase \citep{de2018experimental}. As one of the simplest models, we start with the interacting version of the SSH model which has been recently realized via Rydberg atoms and describes a topological phase transition between a SPT non-trivial and trivial phase.  This model can be described by hard-core bosons $b_{i}^2=0=b_{i}^{\dagger 2}$. We consider two alternating coupling constants denoted by $J$ and $J'$ between adjacent sites $(n,n+1)$, where the former couples sites with odd $n$, and the latter couples sites with even $n$,
\begin{equation}
    H_{J,J'}=\sum_{ij} J_{ij} (b_{i}b_j^{\dagger}+b_{j}b_{i}^{\dagger}),
\end{equation}
As shown in \cite{de2018experimental} this model is mapped to the fermionic SHH model whose energy spectrum consists of two bands
separated by a spectral gap $2(|J|-|J'|)$. Most importantly, this model for $|J'|>|J|$ is a topologically trivial phase while for $|J'|<|J|$ it becomes a topological SPT phase which with open boundary conditions hosts delocalized zero-energy edge
modes. In this case, the symmetry group $G = \mathbb{Z}_2 \times \mathbb{Z}_2$ based on group cohomology classification only has one non-trivial topological phase since $H^2[\mathbb{Z}_2 \times \mathbb{Z}_2;U(1)]=\mathbb{Z}_2.$
Using Jordan-Wigner transformation, the original representation of this model transforms into a spin chain Hamiltonian \cite{de2018experimental}:
\begin{equation}
H=-J'\sum_{i} \sigma_{2i-1}^{-}\sigma_{2i}^{+}-J\sum_{i} \sigma_{2i}^{-}\sigma_{2i+1}^{+} + h.c.
\end{equation}
We use the natural spin basis as our sampling result. We can see in Fig.~\ref{fig:model3}(a) that our method reproduces the phase diagram obtained in \cite{de2018experimental}.

{\it Model 2: $AKLT$.---}
%Next, we continue with more involved SPT models such as the models which realize the Haldane phase in spin-$1$ chains \citep{kennedy1992hidden} with the possibility of hosting several different symmetry broken phases with local order parameters. 
%From model 1 we learn that we can distinguish symmetry broken from symmetry protected phases in a model with a bosonic $U(1)$ symmetry or equivalently fermionic parity symmetry. 
Here, compared to model 1, as a more challenging situation, we ask whether our approach can detect SPT phases with more involved symmetries where symmetry broken phases may exist, too. The Haldane phase of $S = 1$ antiferromagnetic spin chains is a well-known example of such SPT phases with a $\mathbb{Z}_2 \times \mathbb{Z}_2$ symmetry group whose parent Hamiltonian can host several symmetry broken phases. An example of states in the Haldane phase which can be written in a closed form, is the AKLT state \cite{affleck2004rigorous}. This state is a special case of the ground state of the following bilinear-biquadratic spin-$1$ Hamiltonian (${\textstyle {\vec {S_{i}}}}$ are spin-1 operators) when $\theta=1/3$, 
\begin{equation}
\label{eq:bilbiq}
H_{\rm{bb}} = \sum_{i} \cos(\theta_{\rm bb}) (\vec{S}_{i}\cdot \vec{S}_{i+1})
+ \sin(\theta_{\rm bb}) (\vec{S}_{i}\cdot \vec{S}_{i+1})^2.
\end{equation}
However, by varying $\theta$ this model describes two topological and non-topological phase transitions. 
%\begin{figure}
    %\centering
    %\includegraphics[scale=0.30]{model1.pdf}
    %\caption{Detection of phase transitions in model 2.}
    %\label{fig:model1}
%\end{figure}. 

%This Hamiltonian whose first terms is similar to the one-dimensional spin-1 quantum Heisenberg spin model, has an additional ``biquadratic'' spin interaction term which creates a gap above the ground state. 
We can apply the DMA to this model with $N=9$ and plot the resulting phase diagram Fig.~\ref{fig:model3}(b) which is akin to the results in  \cite{gils2013anyonic}. Here, we again see another successful detection of an SPT phase where two additional symmetry broken phases may compete with each other. 

{\it Model 3: Variants of $AKLT$.---}
As an another more complicated model, we study a variant of the $AKLT$ model where in addition to symmetry borken and topological phases we also have a time reversal invariant (TRI) phase with the same symmetries as the Haldane phase \cite{gu2009tensor},
\begin{equation}
    H=\sum_{i}  (\vec{S}_{i}\cdot \vec{S}_{i+1})+ B\sum_{i} S_x^{i}+\sum_{i} D (S_z^{i})^2.
\end{equation}

\begin{figure}
    \centering
    \includegraphics[scale=0.22]{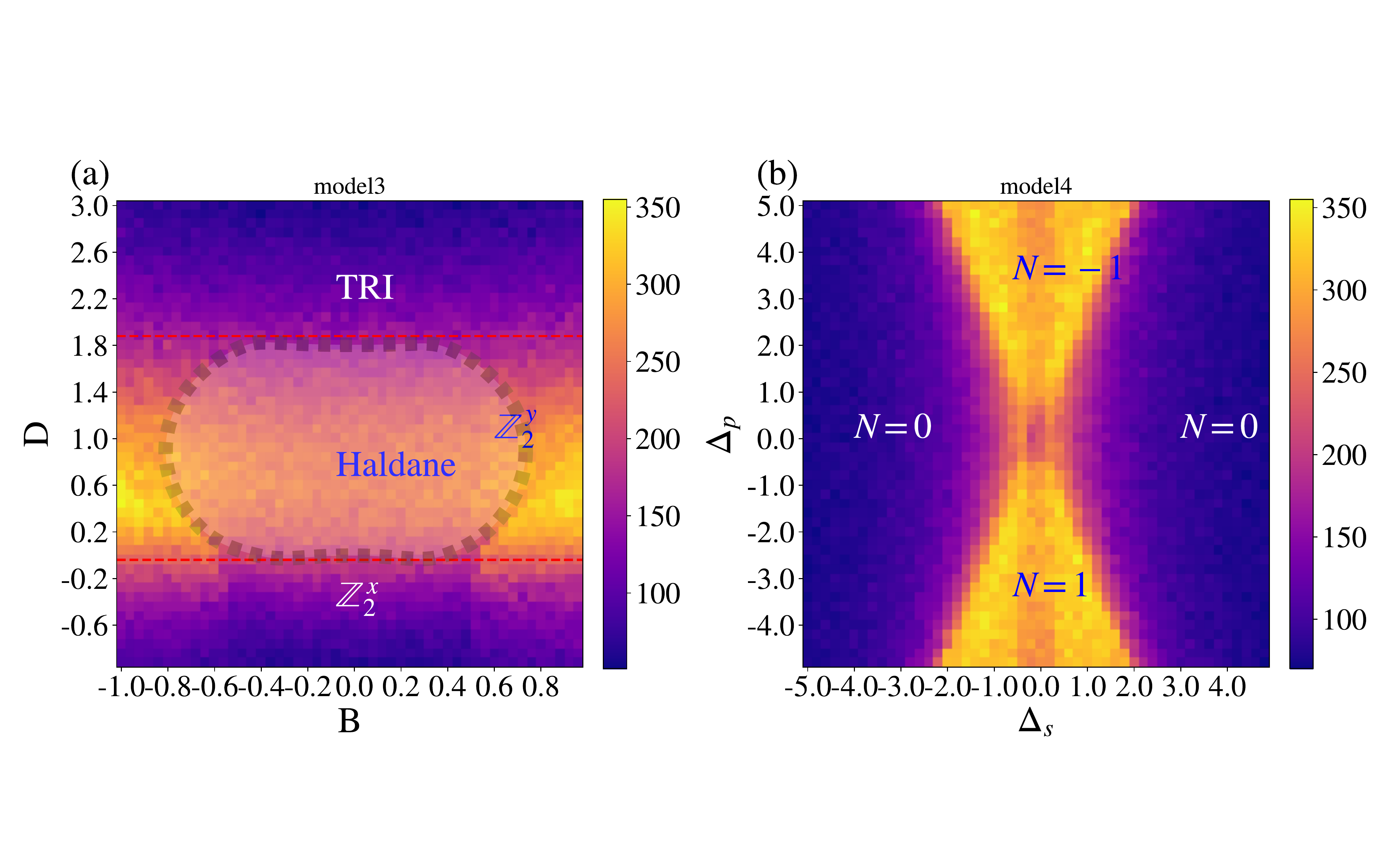}
    \caption{Phase diagrams of model 3 and model 4. (a) Model 3 is a variant of the AKLT model which hosts one trivial phase (TRI), two symmetry breaking phases ($\mathbb{Z}_2^y$, and $\mathbb{Z}_2^x$), and one topologically ordered phase (Haldane). The topological classification of the system is described by the $\mathbb{Z}_2$ group. The dashed line shows the approximate boundaries of the Haldane phase obtained via exact methods. (b) Model 4 is a weakly interacting fermionic model with unconventional pairing. %, whose topological classification is described by the group $\mathbb{Z}_{2}\times \mathbb{Z}_2$. 
      $C$ labels the states via their distinct zero modes in different phases which is obtained via exact diagonalization. Colorbars represent the number of eigenvalues larger than $1-\delta$ (See text).}
    \label{fig:model34}
\end{figure}

This Hamiltonian is invariant under translational symmetry, $\mathbb{Z}_2$ time reversal symmetry $S_y \to -S_y$ and $\mathbb{Z}_2$ parity symmetry. The complete description of the phase diagram is found in \cite{gu2009tensor} which contains a time reversal invariant (TRI) phase with $S^z=0$, and two symmetry broken phases, $\mathbb{Z}_2^{x,y}$, and the Haldane phase. Our results for this model is depicted in Fig.~\ref{fig:model34}(a). In this figure by comparing our phase diagram with those obtained in \cite{gu2009tensor}, we can see a noticeable transition between the $\mathbb{Z}_2^x$, TRI and the Haldane phase. However, the distinction between the $\mathbb{Z}_2^y$ and the Haldane phase is less pronounced and we only see a partial change of color between these two regions. 
%\textcolor{red}{One can build the phase diagram by computing the cluster of each $B, D$ Fig.~\ref{fig:model1}. It can be shown that the diffusion at least $1+1$ D can distinguish the SPT phase and the symmetry broken phase in the periodic boundary condition. This is also convincing since now the on-site symmetry $G$ has been broken. So the state structure can be reflected on our sampled states which explains the efficacy of our method. }

\iffalse
\begin{figure}
    \centering
    \includegraphics[scale=0.15]{model2.pdf}
    \caption{Phase diagram of model 2 in terms of $D, B$}
    \label{fig:model2}
\end{figure}
\fi

{\it Model 4.---}
In models $1-3$ based on group cohomology arguments the topological classification of the phases is described by the $\mathbb{Z}_2$ group, and there is only one topologically non-trivial phase. Here, in addition to spin-$1$ chains, we consider a \textit{weakly} interacting fermionic spin-$1/2$ Hamiltonian which can host more than one non-trivial SPT phase whose topological classification is described by the $\mathbb{Z}_2\times \mathbb{Z}_2$ group. 
%interacting $1$D ``fermionic'' system which cannot be described by the well-known $10$-fold classification of non-interacting fermionic Hamiltonians \citep{Zirnbauer1996Riemannian, Schnyder2008Classification, kitaev2009periodic}. 
This model respects both time reversal symmetry and $S_z$ spin-rotation symmetries. Hence, in the absence of interactions, based on the $10$-fold classification, its topological phases are described by the $\mathbb{Z}$ group. Nevertheless, by adding interactions to this Hamiltonian, the resulting topological classes of this system will reduce to the $\mathbb{Z}_2\times \mathbb{Z}_2$ group. The BdG form of this Hamiltonian is given by,  \citep{Evelyn2012Interacting}:
\bea
 H=-\sum_{<ij>\sigma}c^\dagger_{i\sigma}c_{j\sigma}-2\Delta_s\sum_jc^\dagger_{j\uparrow}c^\dagger_{j\downarrow} \nonumber \\
 \pm i\Delta_p/2\sum_j(c^\dagger_{j+1\uparrow}c^\dagger_{j\downarrow}+c_{j+1\downarrow}c_{j\uparrow})+ \textrm{h.c.},
\eea
%where the first term represents a nearest-neighbor hopping,
%the second term is an on-site pairing, and 
where the last
term is an unconventional pairing between electrons on adjacent sites. This model, can host three phases labeled by $M=0, +1, -1$, which represents the states with distinct zero modes \footnote{As demonstrated in \cite{Evelyn2012Interacting} by stacking two chains with $C=+1$, we get another non-trivial phase with two zero modes. In the presence of strong interactions the phase with $C=-1$ and $C=3$ are identified and hence the classification is $\mathbb{Z}_4$.}. We simulate this model with $N=6$, and plot the resulting phase diagram in Fig.~\ref{fig:model34}(b). 

 \iffalse
\begin{figure}
    \centering
    \includegraphics[scale=0.15]{model6.pdf}
    \caption{phase diagram in terms of $\Delta_p,\Delta_s$}
    \label{fig:model4}
\end{figure}
\fi

As we can in this figure, our approach can regenerate the phase diagram of this model as obtained in \cite{Evelyn2012Interacting} with a high accuracy. To be more specific, we see sharp transitions between the topologically non-trivial phases with $M=\pm 1$ and trivial phases labeled by $M=0$. While in principle, from this plot we cannot distinguish the $M=\pm 1$ phases from each other, from the fact that we see a conspicuous change in the number of eigenvalues at their intersection at $\Delta_{s,p}\simeq 0$ we may speculate that the two regions correspond to distinct phases. This indicates that DMA not only can differentiate between topologically trivial phases and topologically non-trivial phases, but also can provide evidence for differentiating between different topological phases with similar properties.

{\it Discussion and Outlook.---}

In this work, we have used an unsupervised learning algorithm called diffusion maps to detect topological phase transitions between different symmetry broken phases and different SPT phases. %We noticed that once the hyperparameters of the DMA are selected appropriately, this method often shows conspicuous changes in the number of relevant eigenvalues of the diffusion map kernel function which emulates the phase diagram obtained via other methods with an acceptable precision. %For some cases, however, this transition may not be as conspicuous and the boundary of the phases could be smooth. In such situations which 
Our results demonstrate that using diffusion maps as a computationally inexpensive method, can provide helpful evidence for the presence of SPT phase transitions and SPT phases. Since previously this algorithm has been used for detecting other symmetry broken phase transitions in one dimension \cite{lidiak2020unsupervised}, and due to the lack of intrinsic topological order in one dimension, our results demonstrate that the DMA can detect all different types of equilibrium phase transitions in one dimension.

From a conceptual viewpoint, the reason behind the efficacy of the DMA in detecting SPT phases is thought-provoking. One possible explanation is that the DMA in the continuum limit approximates the differential heat equation. On the other hand, the spectrum of elliptic differential operators such as the heat kernel appearing in the heat equation, can include important information about topological invariants  \cite{atiyah_patodi_singer_1975, atiyah_patodi_singer_1975v2}. Based on our results, we speculate that these conceptual relations between the spectrum of differential operators and SPT invariants have practical application for computational and experimental purposes. % and could be theoretically a rich topic for future inspections.   % is a well known fact in mathematics \cite{atiyah1973heat}.  %While we cannot provide a rigorous mathematical proof for this successful application of the DMA, however, 
%Here, based on general grounds we argue that such a successful performance should not be surprising. To elaborate, we recall that diffusion maps in the continuum limit approximate the differential heat equation. On the other hand, the relation between the spectrum of elliptic differential operators such as the heat kernel appearing in the heat equation, and topological invariants of mappings which can be obtained via the Atiyah-Singer index theorems \cite{atiyah_patodi_singer_1975, atiyah_patodi_singer_1975v2} is a well known fact in mathematics \cite{atiyah1973heat}. On the physics side, the relation between the index theorems and bulk-edge correspondence in SPT systems especially the non-interacting ones described by the $10$-fold classification such as the Chern insulators has a long history \cite{Avron1994, bellissard1994noncommutative, hayashi2017bulk}. However, the relation between such differential operators and group cohomology topological invariants pertaining to interacting SPT phases has been more recently pointed out \cite{Witten2016Fermion}. Therefore, our numerical results here provide further evidence that these conceptual relations between the spectrum of differential operators and SPT invariants have practical application for computational and experimental purposes and could be theoretically a rich topic for future inspections. 

An immediate question to be pursued in future is the detection of SPT phases in higher dimensions via the DMA. Another intriguing question to investigate, is the detection of long-range topological order in symmetry enriched topological phases \cite{Mesaros2013Classification} which can be realized in two and higher dimensions without using thermal sampling \cite{scheurer2020unsupervised}. Also, detection of topological phases in non-equilibrium phases of matter especially Floquet systems is left for future studies.

\begin{acknowledgments}
We are grateful to useful discussions with Mohammad Hafezi, Victor Albert, and Alireza Seif for helpful discussions. We acknowledge funding from MURI-AFOSR FA95501610323 and MURI-ARO W911NF-15-1-0397.

\end{acknowledgments}

\bibliography{reference}

\newpage
\clearpage
\widetext

\section*{SUPPLEMENTAL MATERIAL}

In this supplemental, we consider two additional spin chain models which can host SPT phases. We consider the following Hamiltonian with periodic boundary condition as considered for other models in the main text, \citep{kennedy1992hidden}:
\begin{equation}
H=\sum_{i} S_x^{i} S_x^{i+1}+S_y^{i} S_y^{i+1}+\lambda  S_z^{i} S_z^{i+1}+ D (S_z^{i})^2.\label{H1}
\end{equation}
We can use the same techniques and with $L=9$ and $2500$ samples to implement the diffusion map algorithm (DMA). We plot the phase diagram via the DMA. The phase diagram Fig.~\ref{fig:model4} matches with the predictions in \citep{kennedy1992hidden} to a great extent. In particular the distinction between the antiferromagnetic phase (AFM), Haldane and the ferromagnetic phases are completely distinguishable, however, we observe a less evident transition between the Haldane and the dimerized phases. We also notice that we have not detected the $XY$ phase whose presence is still undetermined via exact methods \citep{kennedy1992hidden}.

\begin{figure}
    \centering
    \includegraphics[scale=0.16]{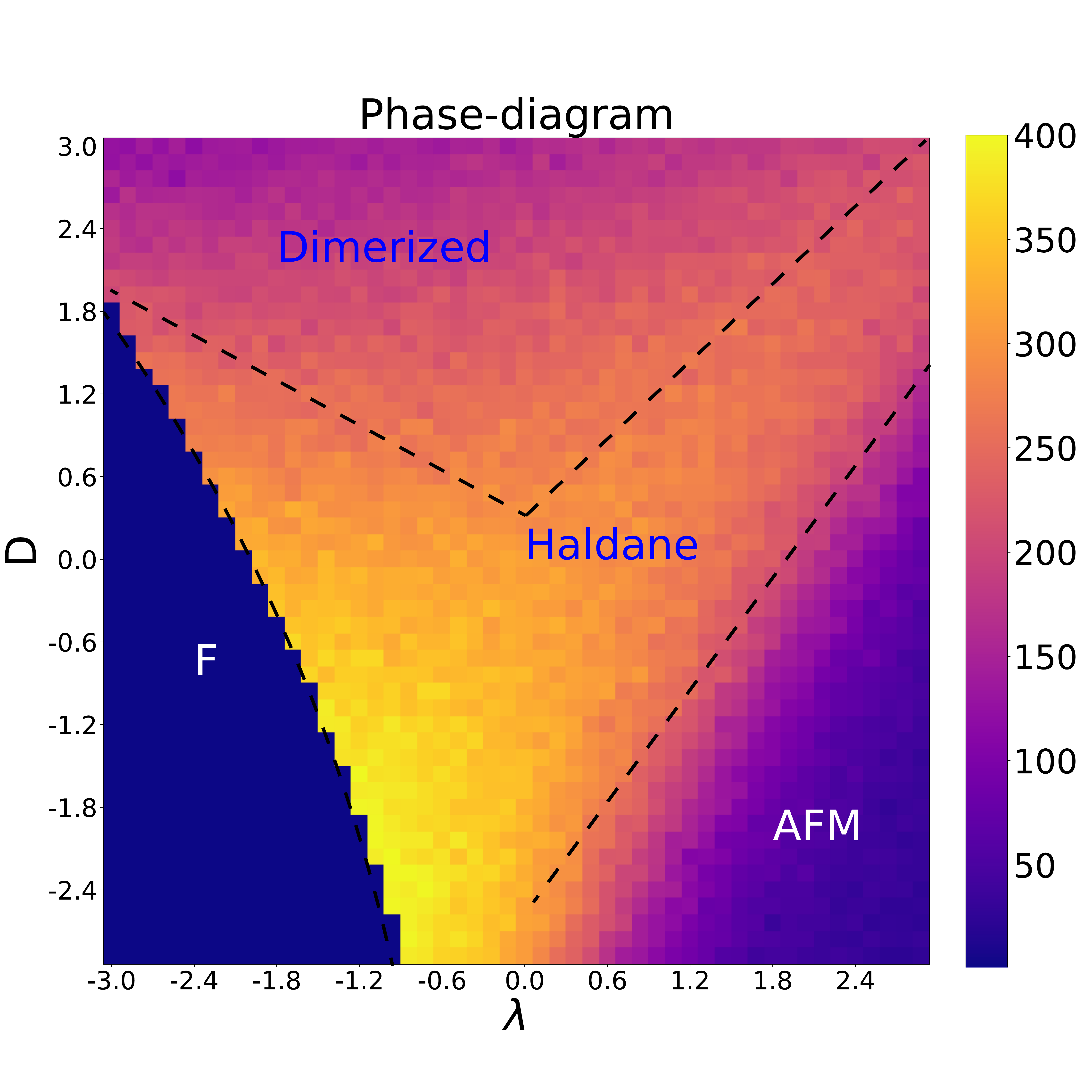}
    \caption{Phase diagram of Hamiltonian in Eq.\ref{H1} in terms of $\lambda,D$ obtained via the diffusion maps algorithm. Dashed lines represent the exact phase boundaries obtained by DMRG methods.}
    \label{fig:model4}
\end{figure}

\begin{figure}
    \centering
    \includegraphics[scale=0.16]{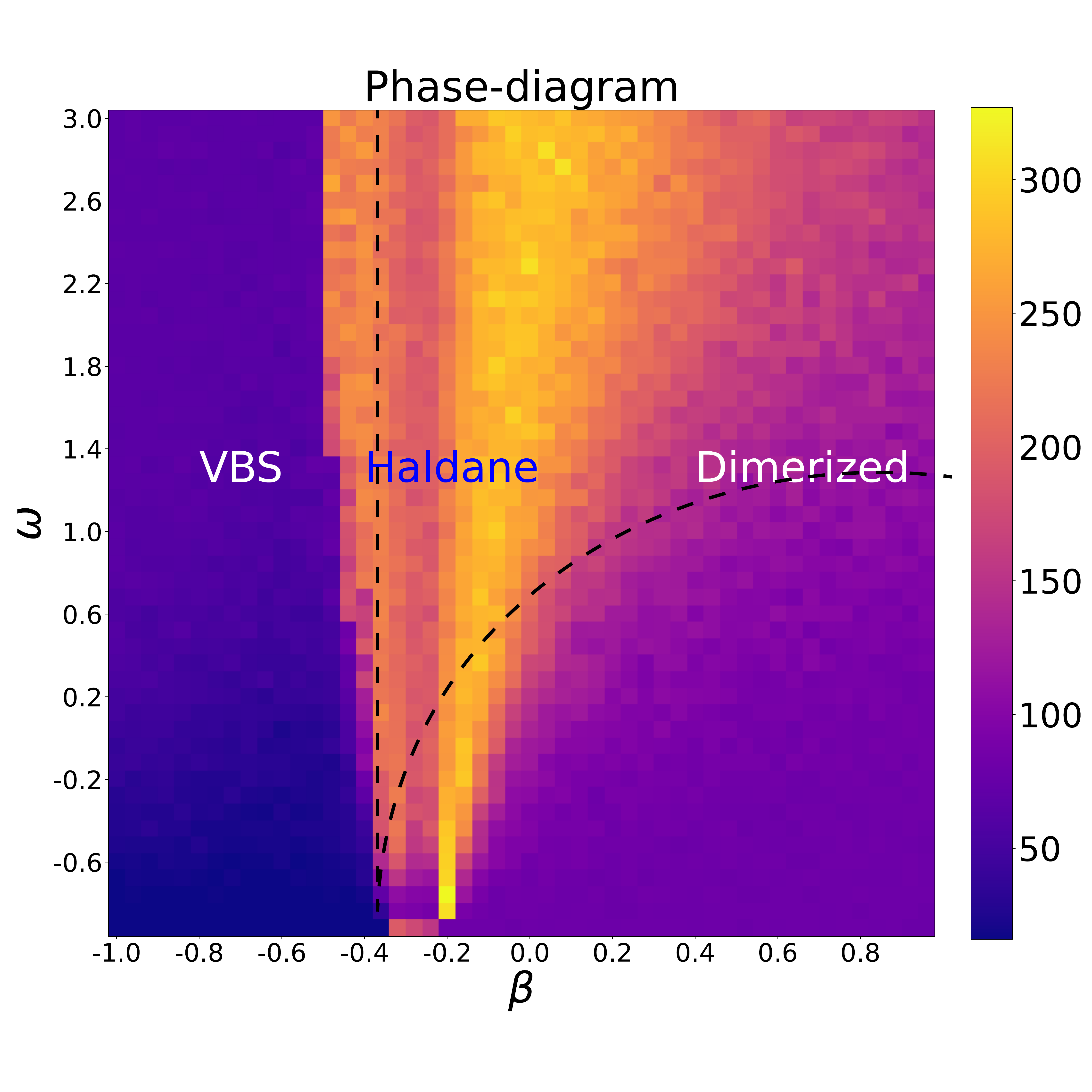}
    \caption{Phase diagram of Hamiltonian in Eq.\ref{H2} in terms of $\beta,\omega$ obtained via the diffusion maps algorithm. Dashed lines represent the exact phase boundaries obtained by DMRG methods.}
    \label{fig:model5}
\end{figure}

The second model that we consider here is the following spin Hamiltonian with periodic boundary condition \citep{kennedy1992hidden}:
\begin{equation}
H=\sum_{i} J_i(S_i \cdot S_{i+1}-\beta (S_i \cdot S_{i+1})^2 ).\label{H2}
\end{equation}
$J_i=1$ for $i$ even, $J_i=\omega$ for $i$ is odd. We can see our method Fig.~\ref{fig:model5} matches qualitatively with \citep{kennedy1992hidden} and we can see a sharp transition between the topological Haldane phase and two symmetry broken phases i.e. valence bond state (VBS) and dimerized phases. We can also predict $\beta=-\frac{1}{3}$ \citep{kennedy1992hidden} for the transition from VBS to Haldane phase. 

\section{PCA and k-means algorithm}
\begin{figure}
    \centering
    \includegraphics[scale=0.16]{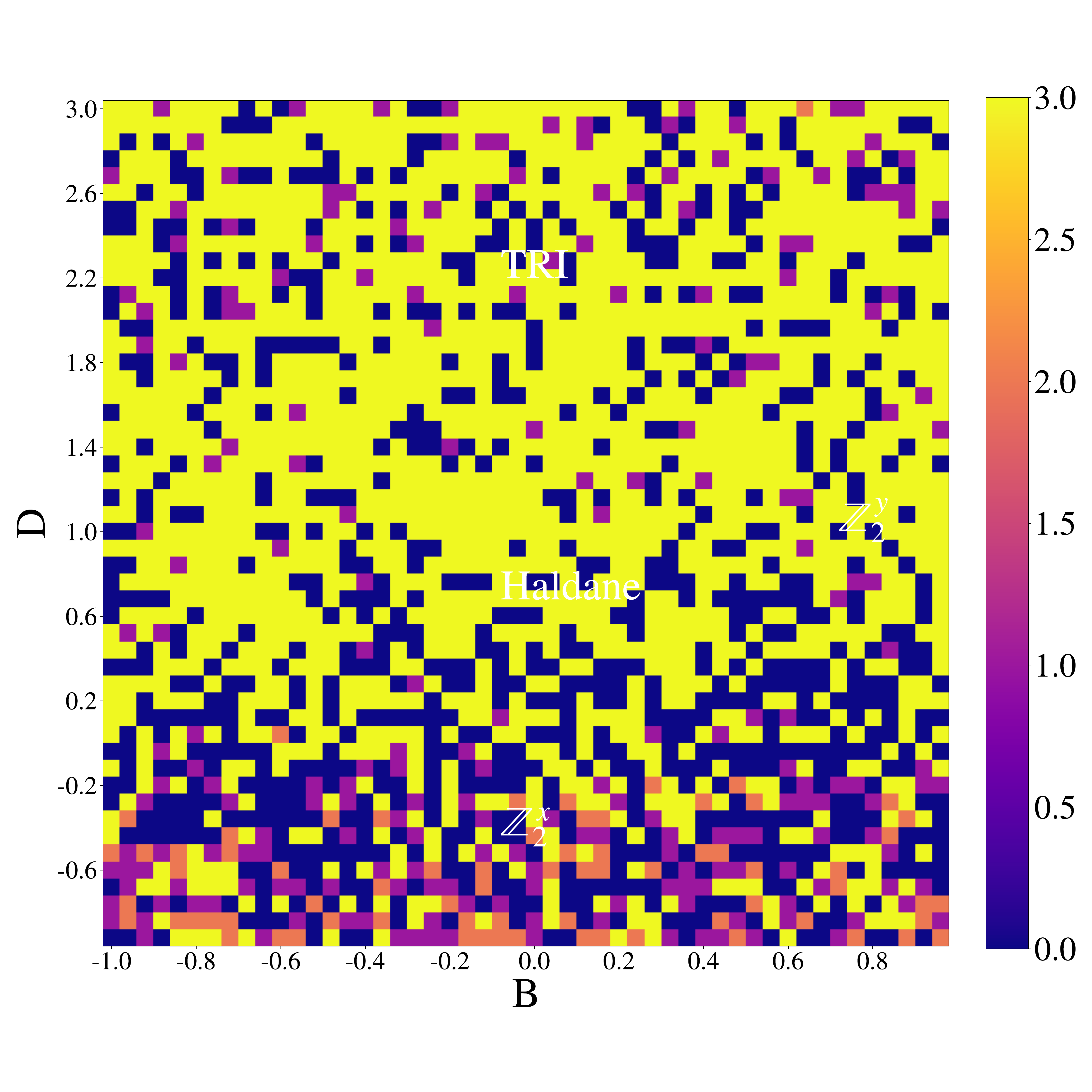}
    \caption{PCA and k means on model 3}
    \label{fig:pca}
\end{figure}

\paragraph{}
To compare the performance of the diffusion maps algorithm and other unsupervised methods, here, we try one of the most standard unsupervised algorithm called PCA (principle component analysis) on our sampling and compute the mean of $2$ components. After doing this, we choose the number of clusters and use the standard $k$ means algorithm. It should be highlighted that we do not have to pick the number of clusters in the diffusion map algorithm. We choose model 3 \cite{gu2009tensor} in our main text.
\begin{equation}
    H=\sum_{i}  (\vec{S}_{i}\cdot \vec{S}_{i+1})+ B\sum_{i} S_x^{i}+\sum_{i} D (S_z^{i})^2.
\end{equation} 
Here, we find that this method fails to identify the phase transition in fig \ref{fig:pca} completely. This superior performance of the diffusion maps in comparison to other unsupervised learning methods, is due to the fact that in this approach we use non-linear kernel functions which allow us to detect complicated non-linear structures in data sets. 

\end{document}